# Brazil Nut Effect in Roads that Allow Cars and Motorbikes to Pass Through

Fisca Dian Utami, Desyana Olenka Margaretta, Donny Dwiputra, Dui Yanto Rahman,

Euis Sustini, and Mikrajuddin Abdullah[a]

Department of Physics

Bandung Institute of Technology

Jl. Ganesa 10 Bandung 40132, Indonesia

[a]Email: mikrajuddin@gmail.com

Abstract

In some countries, cars and motorbikes are allowed to pass through the same road. We observe a phenomenon similar to a Brazil nut effect (BNE) that frequently occuron such roads, in which motorbikes (representations of small particles) segregate in front of cars (representations of large particles), although the motorbikes are initially located far behind the cars. Deep investigations of the BNE-like phenomena in such roads are reported here, where a combination of aerial views using a quad-copter camera and simulationsisstudied. We show for the first time that most phenomena observed in common BNEs are also observed in such roads.

PASC number(s): 02.70.Rr, 89.40.Bb, 89.65.Cd

## I. INTRODUCTION

The BNE appears in various physical-related systems, such as pharmaceuticals, powder metallurgy, glass industries, food, and cosmetics [1], and it is involved in soil layer formation due to seismic shaking as well as size sorting in asteroids [2,3]. Until recently, BNE explorations have mainly focused on systems of small particles (very large numbers) and larger particles (very small numbers) mixtures [4-10], and there are fewer reports on BNE simulations that use a comparable number of small and larger particles [11,12].

The challenging process that likely mimics the granular evolution in a BNE container is traffic flow on most roads in countries where motorbikes (small size) and cars (large size) are allowed to pass. Such roads are easily found in ASEAN countries (Indonesia, Vietnam, Laos, Cambodia), India, Taiwan, and others. We have reasons to claim that this traffic process and granular motion in a BNE container under shaking are fundamentally similar. The main ingredient in the BNE is shaking and mixing of small and larger particles. The energy is supplied from outside, and different grains might receive different amount of energy. The main results of such processes are continuously giving energy to particles and continuous creating voids that located randomly throughout the materials. Once the voids are created just beneath a grain and the grain can occupy the void (the void size is larger than the grain size), the grain suddenly falls to the void, and expressing a downward movement. During shaking, collision between grains occurs and from mechanical point of view, collision means repulsive forcing between grains.

In the case of the above mentioned traffic, we can also define processes that similar to supplying of energy, shaking, and mixing. The in the BNE, the grains receive energy from outside, the magnitude of each might be different, while in the case of traffic, each vehicle produces their own energy, and the magnitude of each might be different. Pressing the accelerator and brake pedals in vehicles might be compared with shaking of granular. Shaking causes granular to move or stop, similar to pressing the accelerator or brake pedal that tend to move or stop the vehicles. The movement of vehicles, especially in crowded situation, is usually irregular, which can be viewed as mixing of the vehicles. During movement of the vehicles, voids are continuously created at random position, analogous to creation of voids between particles in granular matters.

Small voids between cars in the road facilitate the forward penetration of motorbikes. This process can be interpreted as backward movement of the cars relative to the system's center of coordinate, comparable to the upward movement of larger particles in the container. The traffic light can be considered to be the container base, and the continuous supply of energy by the vehicles' motors resembles the continuous energy supply in shaking the granule-filled container. The granules placed in a container tend to move downward due to gravitation, which is comparable to vehicles that tend to move forward. The granular system is dissipative, which readily loses its energy when the external energy supply is terminated due to collisions with other granules [13]; this aspect is similar to vehicles that readily lose their energy (when amotor is suddenly switched off) due to wheel friction with the road surface or by applying a braking force. The forward movement of motorbikes relative to the cars resembles the upward motion of the larger particles relative to the small particles. In a shaking container, small particles move much more easily than large particles; this aspect resembles a traffic system in which motorbikes move much more easily than cars.

We report here a deep investigation of the BNE-like mechanism in traffic as mentioned above. Both aerial observations using a quad-copter camera and computer simulations were employed. Several crowded roads in Bandung city, Indonesia, were selected as targets for the aerial views. Bandung city is a representation of metro cities in developing countries in which both cars and motorbikes are allowed to pass through the same road.

## II. METHODS

**Aerial Views**

We recorded the movements of vehicles starting from road segments far away from a traffic light until they approached the traffic light on February 24, 2018, at *Pahlawan*and *PHHMustofa* streets and on March 11, 2018 at *Simpang Lima* and *PelajarPejuang* streets. Although we selected only four roads, the traffic at the other roads shows similar characteristics,and thus, the selected roads can represent all roads.

**Simulations**

For simulation purposes, we generated two-dimensional lattices of $6 \times L$ cells along the road, where $L$ is the road segment length. One motorbike occupies one cell, while one car occupies four cells ($2 \times 2$ cells),and thus, three cars, or six motorbikes, or their combination can move in parallel. Our system is two-dimensional, and there have also been reports on two-dimensional BNEs, such as by Schnautz et al [14] and Sanders et al [15].

Initially, the motorbikes and the cars are placed randomly in a specific proportion without overlapping. Since the motorbikes move much more easily than the cars, the positions of the motorbikes were first updated, followed by updating the cars' positions. The updating rules for the motorbikes are as follows: (i) *if the space in the direct front of a motorbike is empty, then it moves forward one cell*. (ii) *If the space in front of a motorbike is occupied and one of its front diagonals is empty and the other is occupied, it moves to the empty position*. (iii) *If the two front diagonals are empty, it selects either position randomly*. (iv) *If all three cells in the front are occupied, then the motorbike stays in its position*. The updating rules for cars are the following: (i) *if at least one cell in the direct front of the car is occupied, then the car stays in its position*. (ii) *The car moves forward one cell as long as two cells in the direct front are empty*. The processes are repeated until all motorbikes overtake all cars, which is identical to the condition when large particles in the granular container have reached the top layer. **Figure 1** illustrates the rulesthat govern the motions of the cars and motorbikes.

The length of a cell is slightly longer than the motorbike's length. The length of a common motorbike is approximately 2 m, and for estimation, we assume that the length of a cell is 3 m since space must be left in the front and rear. This arrangement is consistent with the length of a car,which isusuallyapproximately 5 meters, and by giving a space of 0.5 m at the front and 0.5 m at the rear, the space length required for a car is approximately 6 meters, which is in accordance with the length of two cells. The average speed of the vehicles in a crowded road is approximately 40 km/h (11.1 m/s). One time step is equivalent to the time for the vehicle to move one cell (approximately 3 m/11.1 m/s = 0.27 s). By assuming that this time corresponds to a period of shaking the granular container, we obtain that the frequency of updating is approximately 3.7 Hz, which is comparable to the shaking frequency of a granular mixture, such as 13 Hz [16], 7-11.1 Hz [17], 0.5 – 2 Hz for horizontal BNE [14].

One important parameter for describing the BNE is the position of large particles from the container base. In general, the large particles' positions increase linearly with the shaking time and then saturate at a constant height (the granular surface) [2,7,16]. In the traffic case, we will define a parameter that has a similar meaning. Since we are addressing many cars and motorbikes, we defined the cars' center of coordinates and the vehicles' center of coordinates as

$$\bar{X}_c(t) = \frac{\sum_{j=1}^{N_c} X_i(t)}{N_c} \qquad (1)$$

and

$$\overline{X}_v(t) = \frac{\sum_{i=1}^{N_c} X_i(t) + \sum_{j=1}^{N_b} x_j(t)}{N_c + N_b} \qquad (2)$$

respectively, where $X_i(t)$ is the coordinate of the *i*-th car and $x_j(t)$ is the coordinate of the *j*-th motorbike at time *t*, and $N_c$ and $N_b$ are the number of cars and motorbikes, respectively. We then define a quantity that measures the backward movement of the car $\Delta \overline{X}_c(t) = \overline{X}_v(t) - \overline{X}_c(t)$, where $\Delta \overline{X}_c(t)$ is positive when the cars' center of coordinates moves backward relative to the vehicles' center of coordinates. A similar definition for the BNE process has been introduced by Wen et al when considering the segregation of a mixture that contains a comparable number of small and large particles [12].

To understand the flow behavior, we investigate the evolution of the cars' and motorbikes' densities. For this purpose, we marked the road image with several lines that are perpendicular to the road with equal separation. The densities of the cars and motorbikes were calculated using the following rule: (a) *If some part of a motorbike or car is crossed by the line, then we conclude that the motorbike or car exists at that line position*. (b) *If a marked line crosses n motorbikes or cars, we conclude that the number of motorbikes or cars at that line position is n*. (c) *If a marked line does not cross any part of any of the vehicles, the number of vehicles at that line position is zero*.

### III. RESULTS AND DISCUSSION

**Figure 2** shows the time evolution of the car and motorbike densities (the vehicles move to the right), which were extracted from the quad-copter views at (a) *Pahlawan*, (b) *PHH Mustafa*, (c) *Simpang Lima*, and (d) *PelajarPejuang* streets. We extracted the frames at 6 seconds of separation. It is clear from all of the figures that as time evolves, the density of the motorbikes will surpass that of the cars, and the motorbike locations will be at the front. This phenomenon can be considered to be backward motion of the cars, which resembles the upward motion of the larger particles in the shaking granular system.

We also calculated the spatio-temporal motions of the vehicles extracted from the quad-copter views. **Figure 3(a)** shows the trajectories of the cars (yellow) and motorbikes (orange). One curve represents the trajectory of one vehicle. The right horizontal direction indicates the direction of movement, and the vertical direction from the top is the time evolution. It is clear that the forward motion speed of the motorbikes is larger than that of the cars, which leads to segregation.

**Figure 3(b)** is the frames of the quad-copter views separated by 2 seconds. The same vehicles at different times are identified with circles, which also clearly shows that the motorbikes move faster than the car. Finally, **Fig.3(c)** shows that all of the cars were located at the rear positions, which resembles the situation when larger particles in a granular mixture become located at the top.

**Figure 4** is the time evolution of car and motorbike densities based on the simulation. Different ratios of total cars and motorbikes were used: (a) 8 motorbikes+2 cars; (b) 20 motorbikes+20 cars; (c) 100 motorbikes+50 cars; and (d) 240 motorbikes+80 cars. The vehicles' densities were calculated every five time-steps. Initially, the densities of the cars and motorbikes overlapped since all vehicles were placed randomly in the cells. As time evolves, the density of the motorbikes moved faster and finally surpassed the density of the cars and reached the segregated state. It is also clear from **Fig. 4** that the time for segregation depends on the number of vehicles. The times for segregation at conditions (a), (b), (c), and (d) were approximately 20 steps ($\approx$ 5.4 s), 68 steps ($\approx$ 16 s), 200 steps ($\approx$ 54 s), and 320 steps ($\approx$ 86.5 s), respectively.

In contrast, clear segregation was not observed in real traffic based on the quad-copter view since the true road is an open system. Some vehicles left the road, and new vehicles entered the road at random points along the road. The closed road system is found in the freeway, but in this road, the motorbikes are not allowed to pass through, and thus, the BNE will not be observed. However, based on the trajectory in **Fig. 3(a),** the trend for segregation is clearly observed. In some situations, segregation is observed, as shown in **Fig. 3(d).**

The trajectories of the cars and motorbikes from the simulation for 20 cars+20 motorbikes are shown in **Fig. 5(a).** Again, initially, the positions of the vehicles are random (top left). As time evolves, the motorbikes move faster to the right and finally surpass the vehicles and segregate after approximately 68 time-steps ($\approx$ 16 s).

We can further extract from **Figure 4** the evolution of the fraction of motorbikes that have surpassed the cars. **Figure 5(b)** shows such a fraction from a simulation that had 20 motorbikes+20 cars. It is likely that the fraction of motorbikes that have overtaken the cars is initially zero for a certain time period and then increases linearly with time. After approximately 60 time steps, all of the motorbikes have overtaken the car, as is also represented in **Figure 4(c)** or **Fig 5(a).**

**Figure 6** shows the position of the cars' center of coordinate relative to the vehicles' center of coordinate based on the simulation. The positive sign means that the cars' center of coordinate is located behind the vehicles' center of coordinate. Three car and motorbike mixtures were tested: (motorbikes number, car number) = (20,20), (50,100), and (80,240). Initially, the cars' center of coordinate and the vehicles' center of coordinate coincided or nearly coincided, which is in accordance with a random placement of the vehicles. In Fig. 5(a), the random placement of the vehicles has led to the cars' center of coordinate being slightly at the front of the vehicles' center of coordinate (most motorbikes are located far at the rear). As time elapsed, the distance of the cars' center of coordinate to the vehicles' center of coordinate increases linearly until it reaches the maximum value. By further increasing the time step, the distance decreases since the motorbikes have reached the traffic light and they stop there while the cars are still moving forward. Finally, after all cars also stop due to the traffic light, the distance of the cars' center of coordinate remains constant.

The peak of the cars' center of coordinate was found at different updating steps, depending on the numbers of cars and motorbikes. From **Fig. 6**, the time steps for reaching the peak are between 90 and 300 steps. Since each step corresponds to an interval of approximately 0.27 s, the time durations to reach the peak are in the range of 24-81 s. This time range is very close to the time for the larger particles to reach the top for BNE systems, such as approximately 60 s [14]. The time to reach the peak also increases with the number of vehicles.

The characteristics shown in **Fig. 5** are also observed in BNE systems where the distance of the large particles from the container base increases linearly with the shaking time [8,16]. A nearly exponential rise that can still be well fitted with a linear function was also reported by Sun et al using a simulation of 7,600 small particles and one large particle [7]. Especially when the ratio of diameters of large particles to small particles is large enough (> 4), the large particles stay at the top or experience only a small oscillation [8], while for a diameter ratio of < 4, the large particles that have reached the top layer can penetrate again deeply into the small particles and then rise again repeatedly [8]. In a real traffic situation, the cars are still chasing motorcycles that have stopped early at the traffic light. However, indeed, the two mechanisms are precisely equal if there is no traffic light.

## IV. CONCLUSIONS

BNE-like phenomena have been proven to occur in roads that allow cars and motorbikes to pass through the same road. In that road, the number of motorbikes is usually much larger that the number of cars, which precisely resembles the common BNE system,which has a container that contains a large number of small particles and a very small number of large particles. The motorbikes well represent the small particles, and the cars represent the large particles (intruder). Shaking the container means a continuous supply of energy to the granules,which is mimicked by the continuous supply of energy from the vehicles'motors, andthe dissipation behavior of the granularity is similar to the loss of vehicle energy by friction with the road or by applying a braking force. Both aerial observation using a quad-copter camera and simulations have shown the true BNE effect, including the backward motion of the cars' center of coordinate and the segregation of the motorbikes.

## ACKNOWLEDGEMENT

This work was supported by PMDSU fellowship from The Ministry of Research, Technology and Higher Education, Republic of Indonesia No. 328/SP2H/LT/DPRM/II/2016.

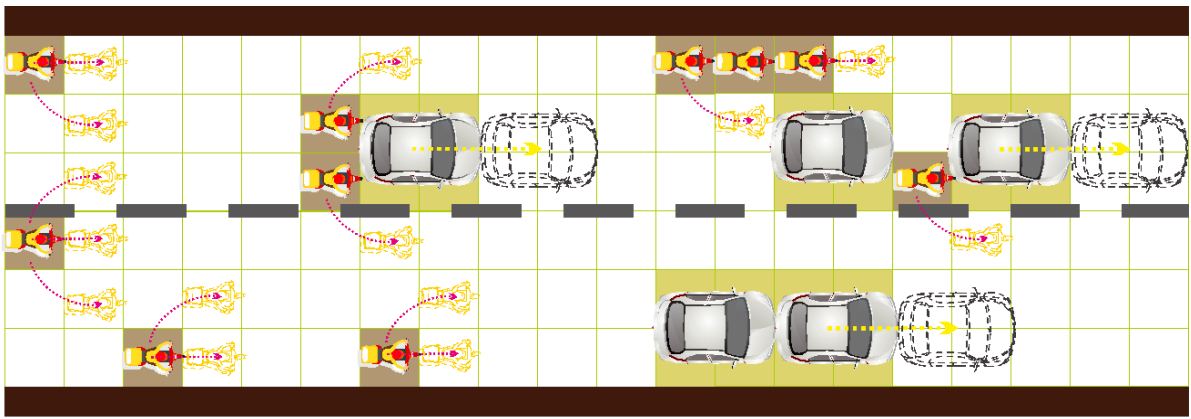

**Figure 1** Cells governed on the road. The road segment is divided into 6×L cells. One motorbike can occupy one cell, while a car can occupy only four cells (2×2 cells). The arrows indicate the allowed movement directions of the vehicles

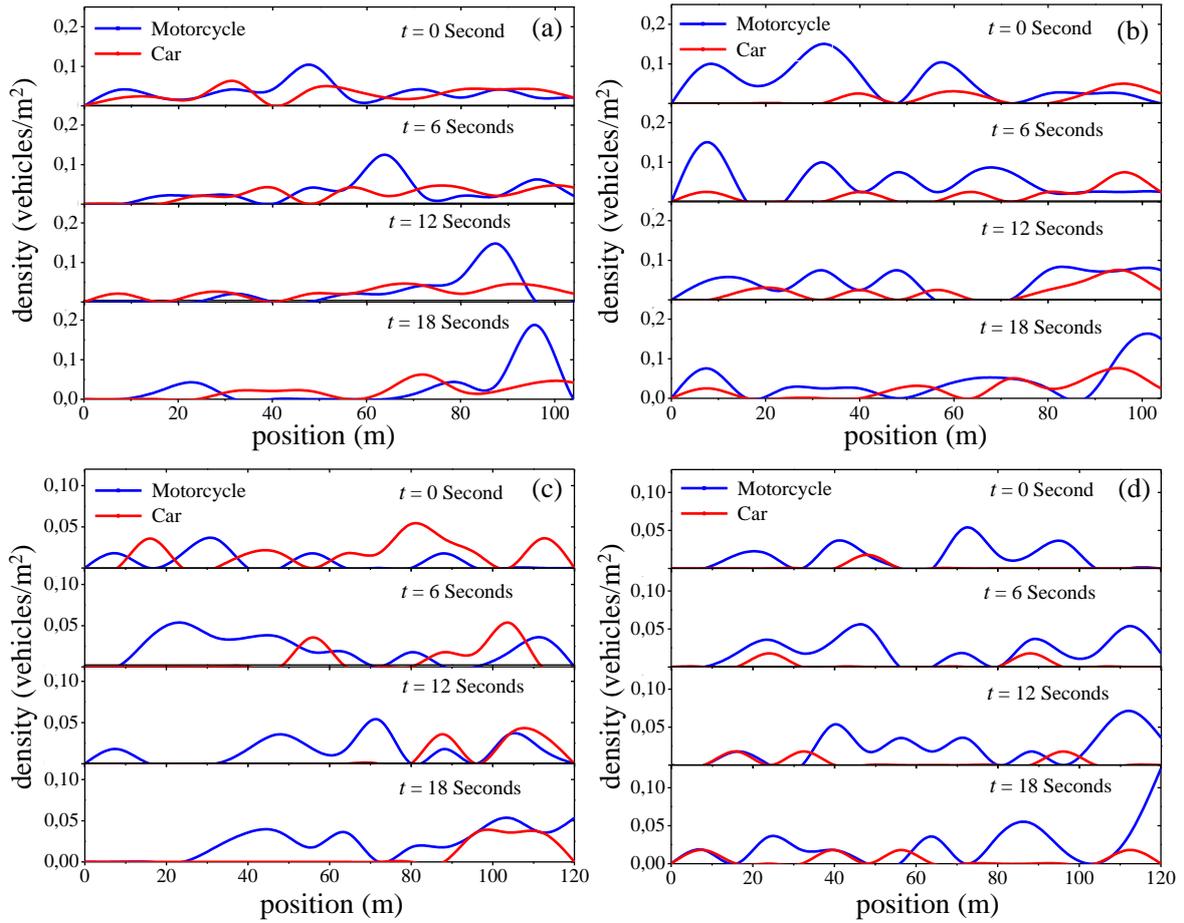

**Figure 2** Time evolution of the car and motorbike densities based on the quad-copter views at (a) *Pahlawan*, (b) *PHH. Mustofa*, (c) *SimpangLima*, and (d) *PelajarPejuang* streets. The blue curves are for the motorbikes, and the red curves are for the cars. The vehicles move from left to right.

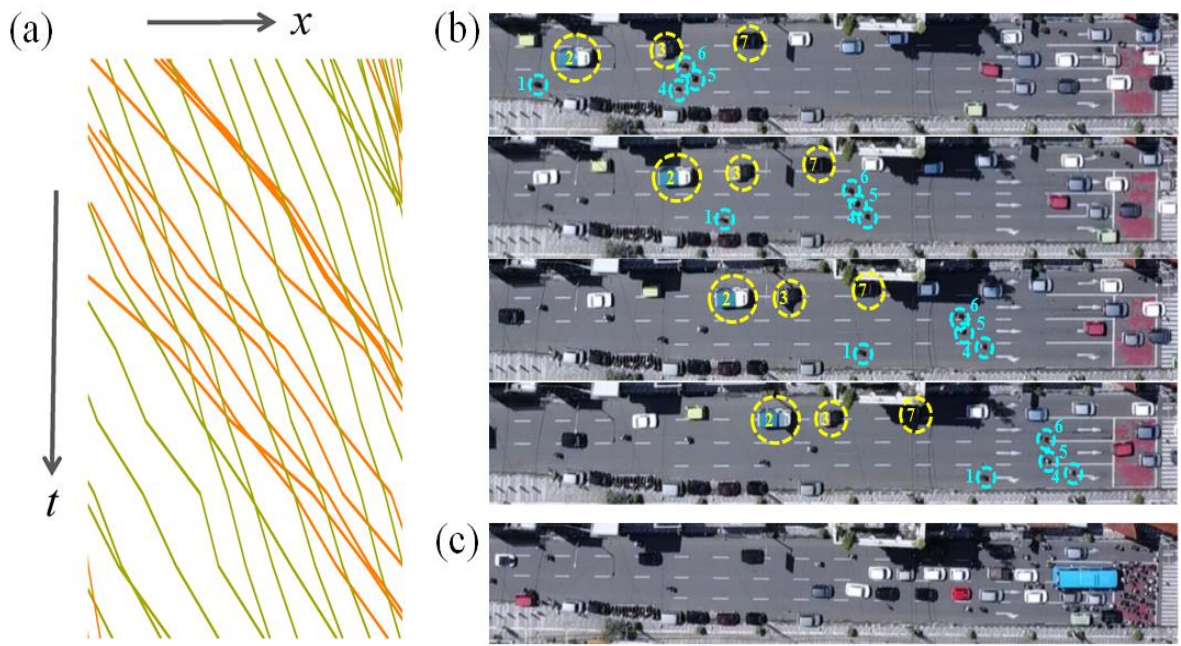

**Figure 3** (a) Trajectory of the cars and motorbikes at *Pahlawan* street extracted from quad-copter views: (orange) motorbikes (orange) and (green) cars. (b) Selected frames from the quad-copter view at *Pahlawan* street for 4 consecutive times separated at 2 seconds. (c) Example of the segregation state where almost all of the motorbikes are located at the front of the cars.

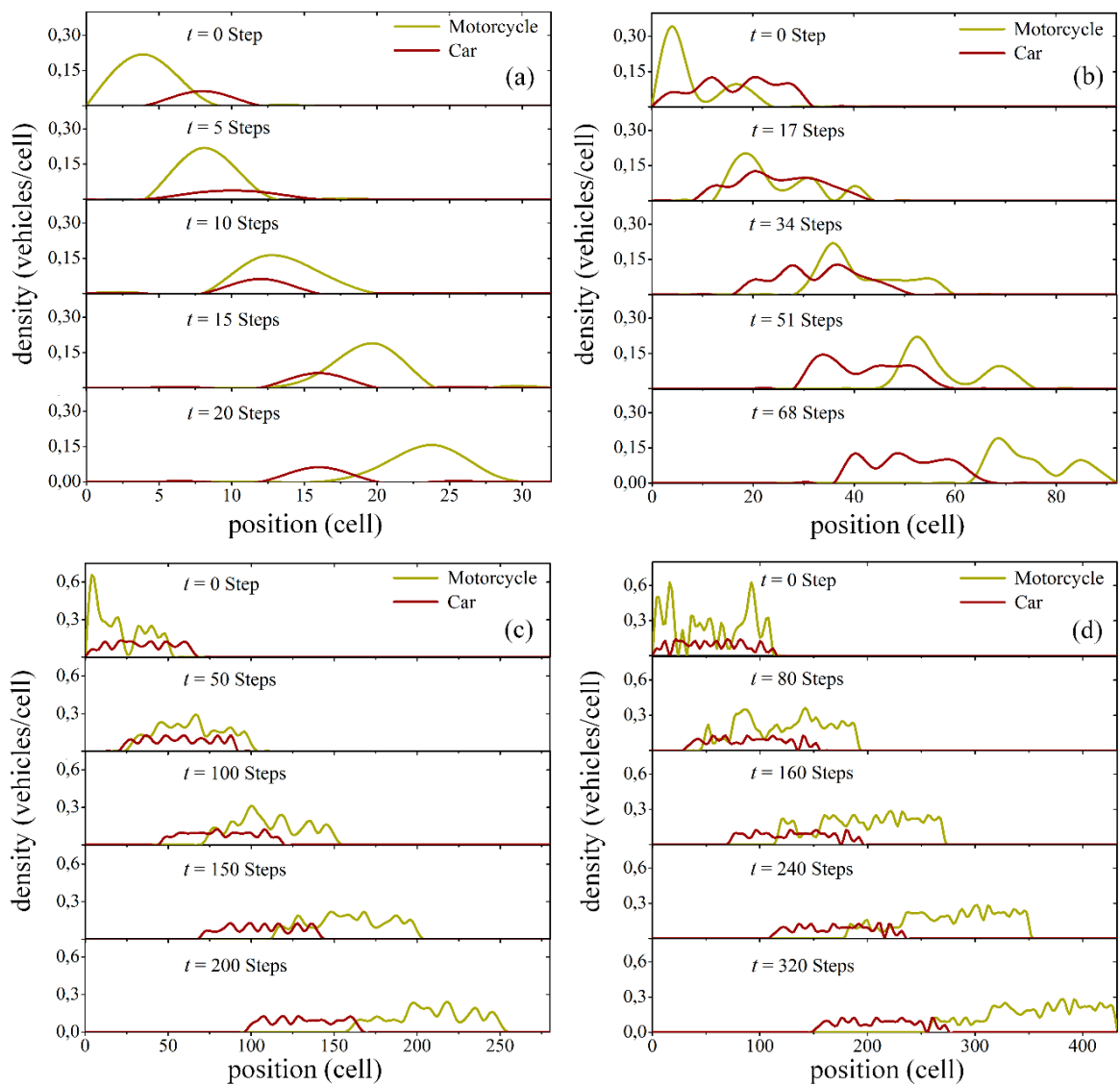

**Figure 4** Time evolution of car and motorbike densities based on simulations with different numbers of motorcycles and cars: (a) 8 motorbikes+2 cars; (b) 20 motorbikes+20 cars; (c) 100 motorbikes+50 cars; and (d) 240 motorbikes+80 cars.

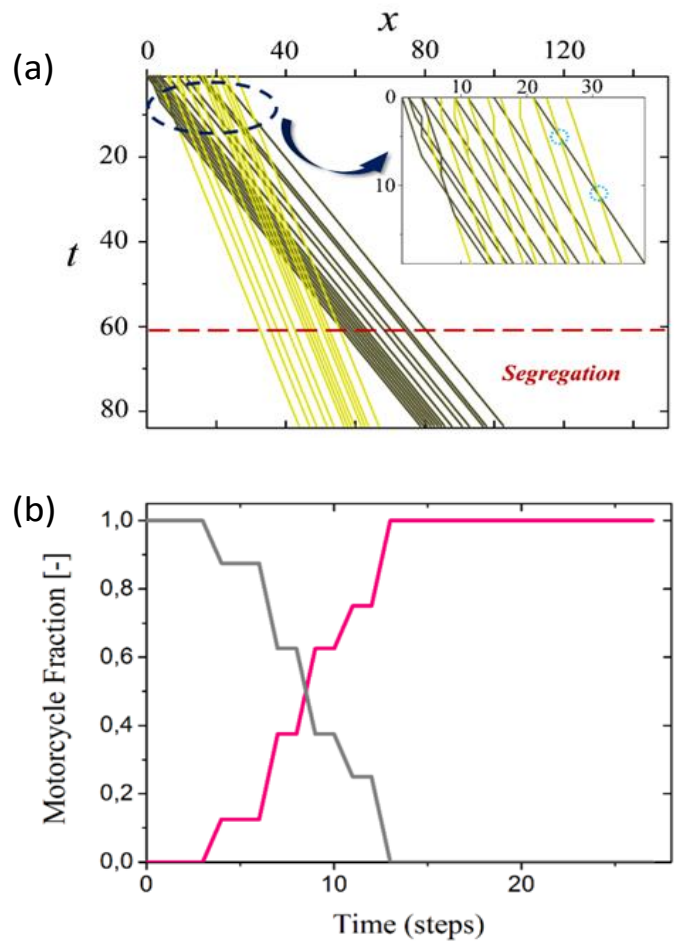

**Figure 5 (a)** Trajectories of cars and motorbikes from a simulation that uses 20 motorbikes+20 cars: (dark grey) motorbikes and (yellow) cars. The vehicles move to the horizontal right, and vertical downward is the time steps. The blue dotted circles at the inset are the condition when the motorbikes are ready to overtake the cars. **(b)** (pink curve) The fraction of motorbikes that have overtaken the cars and (grey curve) the fraction of motorbikes that have not overtaken the cars.

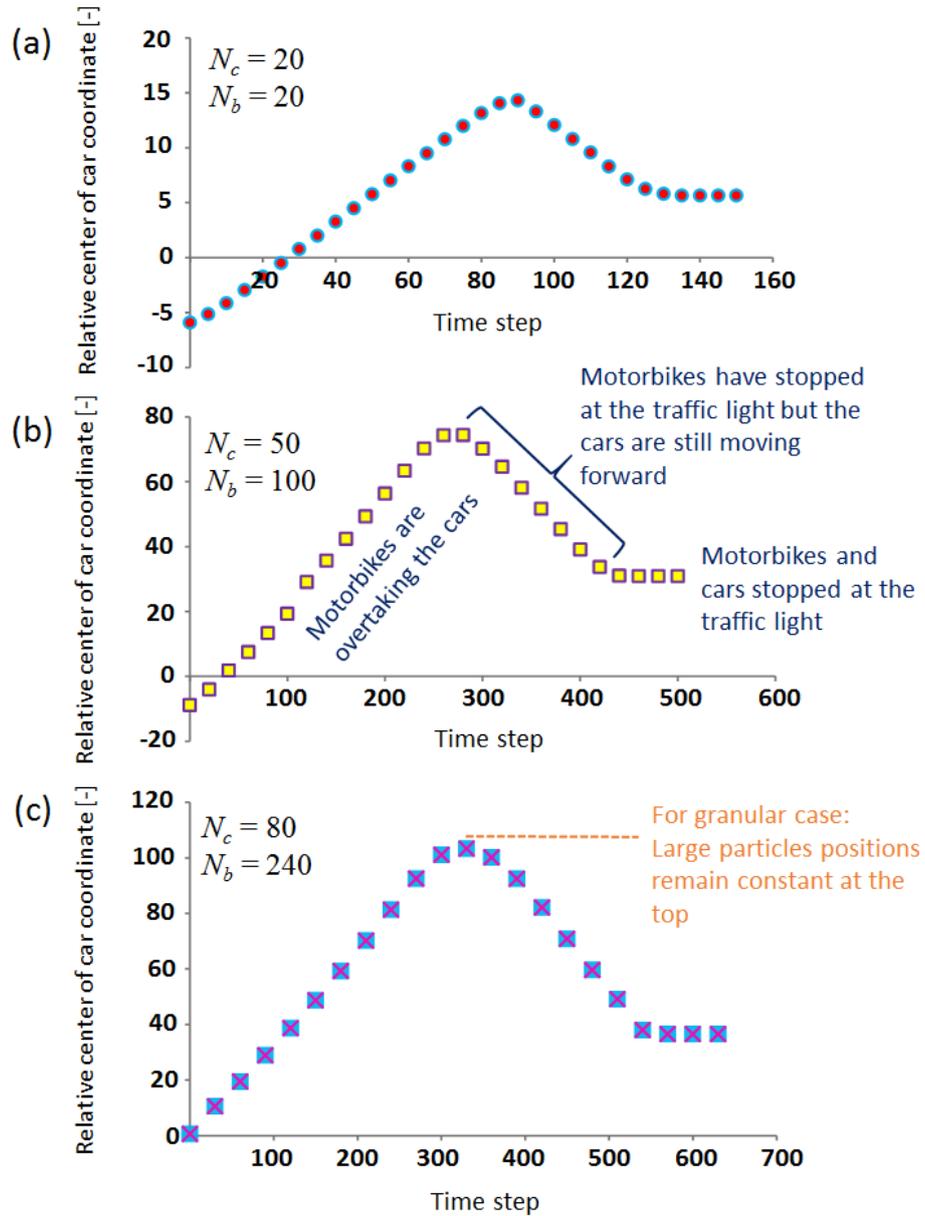

**Figure 6** Evolution of cars' center of coordinates relative to vehicles' center of coordinates: (a) 20 motorbikes+20 cars, (b) 100 motorbikes+50 cars, and (c) 240 motorbikes+80 cars.